\documentclass[aps,pra,floatfix]{revtex4}

\pdfoutput=1

\usepackage{times}
\usepackage{amsfonts}
\usepackage{amssymb}
\usepackage{amsmath}
\usepackage{graphicx}
\usepackage{bm}
\usepackage{verbatim}
\usepackage[cspex,bbgreekl]{mathbbol}
\allowdisplaybreaks

\usepackage{hyperref}

\usepackage{bm} 
\usepackage{color}
\usepackage{stackrel}
\usepackage{accents}

\usepackage{latexsym}

\newcommand{\beg}{\begin{equation}}
\newcommand{\en}{\end{equation}}
\newcommand{\begs}{\begin{subequations}}
\newcommand{\ens}{\end{subequations}}
\newcommand \bea {\begin{eqnarray}}
\newcommand \eea {\end{eqnarray}}
\newcommand{\bem}{\begin{bmatrix}}
\newcommand{\enm}{\end{bmatrix}}
\newcommand{\bpm}{\begin{pmatrix}}
\newcommand{\epm}{\end{pmatrix}}
\newcommand{\bvm}{\begin{vmatrix}}
\newcommand{\evm}{\end{vmatrix}}
\newcommand{\ba}{\begin{array}}
\newcommand{\ea}{\end{array}}
\def\i{\imath}
\def\g{\gamma}
\def\d{\delta}

\def\a{\alpha}
\def\b{\beta}
\def\f{\phi}
\def\ve{\varepsilon}
\def\s{\sigma}
\def\m{\mu}
\newcommand{\by}{\times}

\def\ra{\rightarrow}

\def\inv#1{\frac{1}{#1}}
\def\proj#1{|#1\rangle \langle #1|}
\def\out#1#2{|#1\rangle \langle #2|}
\newcommand{\wt}[1]{\widetilde{#1}}
\def\dis{\displaystyle}

\def\dag{\dagger}
\newcommand{\ket}[1]{|#1\rangle}
\newcommand{\bra}[1]{\langle#1|}
\newcommand{\del}[1]{\frac{\partial #1}{\partial t}}
\newcommand{\re}[1]{(\ref{#1})}

\newcommand{\eref}[1]{Eq.~(\ref{#1})}

\newcommand{\esref}[1]{Eqs.~(\ref{#1})}

\begin{document}

\title{Quantum integrability in the multistate Landau-Zener problem}

\author{A. Patra$^1$ and E. A. Yuzbashyan$^1$}
\affiliation{$^1$Center for Materials Theory, Rutgers University, Piscataway, NJ 08854, USA}

\begin{abstract} 
We analyze Hamiltonians linear in the time  variable for which the multistate Landau-Zener problem  is known to have an exact solution. We show that they either belong to families of
mutually commuting  Hamiltonians polynomial in time or  reduce to the $2\by 2$ Landau-Zener problem, which is considered trivially integrable. The former category includes the equal slope, bow-tie, and generalized bow-tie models. For each of these models we explicitly construct the corresponding  families of  commuting matrices.  The equal slope model is a member of an integrable family that consists of the maximum possible number (for a given matrix size) of commuting matrices linear in time. The bow-tie model belongs to a previously unknown, similarly maximal family of quadratic commuting matrices. We thus conjecture that quantum integrability understood as the existence of nontrivial parameter-dependent commuting partners is a necessary condition for the Landau-Zener solvability. Descendants of the  $2\by 2$ Landau-Zener Hamiltonian are e.g. general $SU(2)$ and $SU(1,1)$ Hamiltonians, time-dependent linear chain, linear, nonlinear, and double  oscillators. We explicitly obtain solutions to all these Landau-Zener problems  from the $2\by 2$ case.

\end{abstract}

\maketitle


\section{Introduction}

Multistate Landau-Zener (LZ) problem is an archetypical problem in non-equilibrium physics, which arises in various experimental setups, see e.g. \cite{Oliver, Altland, Alt, Masing, Petta}. The most frequently encountered scenario deals with a Hamiltonian linear in time and the corresponding non-stationary Schr\"{o}dinger equation      
\beg
\ba{l}
\dis H(t) = A + Bt,\\
\dis \i \del{\Psi (t)} = H(t)\Psi (t),
\ea
\label{LZH}
\en
where $A$ and $B$ are two constant Hermitian $N \by N$ matrices. Given the state of the system at $t \ra -\infty$, the problem is to determine its state at $t \ra +\infty$, i.e. the full asymptotic $S$ matrix containing necessary information about probabilities for all possible transitions. The $2\by 2$ problem was solved  by Landau\citep{Landau}, Zener\citep{Zener}, Majorana\citep{Majorana} and St\"{u}ckelberg\citep{Stuckelberg} in 1932; whereas for dimensions $N \geq 3$, the solution is known only for a few special cases, where $A$ and $B$ have  very specific forms\cite{Demkov1, Brundobler, Ostrovsky, Demkov2}.  

In this paper we attempted to catalog all known exactly solvable multistate Landau-Zener Hamiltonians. We find that they can be classified into two broad categories. The first type of Hamiltonians have nontrivial commuting partners polynomial in $t$, which we explicitly delineate here. We categorize these Hamiltonians  as \textit{quantum integrable}. These include equal slope, bow-tie, and generalized bow-tie models.  As a byproduct of this study we construct a new family of integrable matrices quadratic in $t$. 
Then we have another type of Hamiltonians for which the LZ problem reduces to the $2 \by 2$ case.  These  \textit{descendants} of the $2 \by 2$ LZ Hamiltonian as well as  the $2 \by 2$ LZ  Hamiltonian itself are considered to be \textit{trivially integrable} as we explain below.  For example, we derive solutions of the following LZ problems from the $2\by 2$ one: general $SU(2)$ and $SU(1,1)$ LZ Hamiltonians, time-dependent linear chain, linear oscillator, as well as double and nonlinear oscillators (see below). Similar to the $2 \by 2$ case,  other integrable LZ Hamiltonians produce hierarchies of  solvable descendants through the same procedure. Thus we conjecture \textit{quantum integrability is a necessary condition for Landau-Zener Solvability.}  Integrability therefore  can be used as a preliminary test for identifying new LZ solvable Hamiltonians.  We must clarify that a sufficient condition for LZ solvability is still unknown and  might be more restrictive, i.e.  contain additional requirements besides integrability.

Unlike classical integrability the quantum counterpart is a nebulous concept, where different criteria are commonly used and often it is not  easy to correlate any two of them \cite{Caux2011}. Here we adopt the approach of  \cite{Shastry2002,Shastry2005,Emil1, Shastry2011,Emil2, Shastry} and define a Hermitian matrix  $H(t)$ that depends linearly on a parameter $t$ (e.g. time, interaction strength, external field etc.)  to be  integrable if it has at least one nontrivial commuting partner    polynomial in $t$. Suppose $[I(t), H(t)]=0$ and $I(t)$ is a matrix polynomial in $t$ of order $p$. We say that $I(t)$ is \textit{nontrivial} if it cannot be expressed in terms of powers of $H(t)$ no higher then $p$, i.e. $I(t)\ne\sum_{k=0}^p c_k(t) H^k(t)$, where $c_k(t)$ are (scalar)  polynomials in $t$. From Eq.~\eqref{LZH} we see that the Hamiltonian in the multistate LZ problem  is linear in $t$ and so this definition for quantum integrability is quite appropriate for our purpose. In our paper for each of the  exactly solvable LZ models we stop after finding   nontrivial polynomial commuting partner of the lowest possible order and  we never had to go beyond quadratic parameter dependence. This however does not rule out existence of even higher order nontrivial polynomial commuting partners. This definition of quantum integrability has also been linked with other  well known hallmarks of quantum integrability, such as Yang-Baxter equation, Poisson level statistics and energy level crossings \cite{Shastry}. Let us also note that $[I(t), H(t)]=0$ does not imply conservation of $I(t)$ in time due to its explicit time dependence.

We start with the equal slope model. The Hamiltonian in $N$ dimension is \cite{Demkov1,Brundobler}
\beg
H(t) =\bem
0 & p_{2} & \cdots & p_{N}\\ 
p_{2}^*& a_{2} & \cdots & 0\\ 
\vdots & \vdots & \ddots &  \vdots\\ 
p_{N}^* & 0 & \cdots & a_{N}
\enm + t\bem
b & 0 & \cdots & 0\\ 
0& 0 & \cdots & 0\\ 
\vdots & \vdots & \ddots &  \vdots\\ 
0 & 0 & \cdots & 0
\enm .
\label{eqsl}
\en
 Here and in bow-tie and generalized bow-tie models below we assume $p_i\ne 0$ and $r_i\ne r_j$ for all $i\ne j$.  We show that this Hamiltonian has maximum possible number of linearly independent commuting partners linear in $t$  and thus belongs to the \textit{maximal} family of commuting Hamiltonians linear in a parameter constructed in Refs.~\cite{Shastry2005,Emil1}. 

 Another finite dimensional LZ Hamiltonian that can be solved is the bow-tie model \cite{Ostrovsky} 
 \beg      
H(t)=\bem
0 & p_{2} & \cdots & p_{N}\\ 
p_{2}^*& 0 & \cdots & 0\\ 
\vdots & \vdots & \ddots &  \vdots\\ 
p_{N}^* & 0 & \cdots & 0
\enm + t\bem
0 & 0 & \cdots & 0\\ 
0& r_{2} & \cdots & 0\\ 
\vdots & \vdots & \ddots &  \vdots\\ 
0 & 0 & \cdots & r_{N}
\enm .
\label{bt}
\en
We show that this Hamiltonian does not possess any nontrivial linear commuting partners but  has quadratic ones for dimensions $4 \by 4$ and above. In this paper we explicitly write down the general quadratic commuting partner. By counting the number of independent parameters one can show that the number of linearly independent quadratic commuting partners equals to $N$,  the dimension of the Hamiltonian. This number includes multiples of $H^2(t)$, $H(t)$, and the identity matrix and   therefore in the $3\by 3$ case  nontrivial quadratic commuting partners are absent. We  determine all quadratic   matrices that commute with the bow-tie Hamiltonian and each other, thus  constructing an integrable family with only one linear and the  maximum number of quadratic commuting matrices.        

Demkov and Ostrovsky \cite{Demkov2} considered the following   generalized version of the bow-tie model:  
\beg      
H(t)=\bem
\frac{\varepsilon}{2} & 0 & p_{3} & \cdots & p_{N}\\ 
0 & -\frac{\varepsilon}{2} & p_{3} & \cdots & p_{N}\\ 
p_{3}^* & p_{3}^* & 0 & \cdots & 0\\
\vdots & \vdots & \vdots & \ddots &  \vdots\\ 
p_{N}^* & p_{N}^* & 0 & \cdots & 0
\enm +  t \bem
0 & 0 & 0 & \cdots & 0\\ 
0 & 0 & 0 & \cdots & 0\\ 
0& 0 & r_{3} & \cdots & 0\\ 
\vdots & \vdots & \vdots & \ddots & \vdots\\ 
0 & 0 & 0 & \cdots & r_{N}
\enm .
\label{gbt}
\en
Unlike the bow-tie case, where only one level interacts with the rest of the levels; in the generalized bow-tie case two levels interact with the rest. In both bow-tie and generalized bow-tie case the rest of the levels do not interact. One must also notice that in the case of generalized bow-tie Hamiltonian the two special levels do not interact with each other either. We show that this Hamiltonian possesses a nontrivial linear commuting partner, which we explicitly construct. In addition, using the counting technique   mentioned earlier we show that there can only be one nontrivial linear commuting partner. Hence we call this to be a member of the \textit{minimal} linear commuting family.

Next, we turn our attention to other solvable models. It can be shown that a general $2 \by 2$ LZ Hamiltonian can be written as  $H(t) = \frac{g}{2}\s_{x} + \frac{t}{2}\s_{z}$ by adding a multiple of identity and performing a $t$-independent unitary transformation, where $\s_{x}, \s_{y}$ and $\s_{z}$ are the three Pauli matrices.  These transformations  affect only the overall phase of  $\Psi(t)$ and so they have no effect on the transition probabilities \cite{Brundobler}. This case was originally considered and  solved explicitly by Landau and Zener. However, if we consider the  vector space   of $2 \by 2$    matrices that commute with $H(t)$ and are linear in $t$, the only possible linearly independent elements are the Hamiltonian and multiples of the identity.   Moreover, in Section II we present a general argument proving that a nontrivial polynomial commuting partner of degree $p$ will only come to exist from dimension $(p+2) \by (p+2)$ onward. Evidently then in the $2\by 2$ case  nontrivial   commuting partners  are impossible and so we claim this case to be trivially integrable. Similarly, the $3 \by 3$ bow-tie Hamiltonian in addition to having no linear in $t$ commuting partners cannot also have quadratic ones, unlike  bow-tie Hamiltonians of higher dimensions.    Thus we consider it to be the trivial member among  linear Hamiltonians with  quadratic commuting partners. 

One can  generalize the $2 \by 2$ LZ problem by considering the Hamiltonian $H(t) = gS_{x} + tS_{z}$, where $S_{x}, S_{y}$, and $S_{z}$ are the $x$, $y$, and $z$ components of a general $SU(2)$ spin operator. The time evolution operator for  this Hamiltonian can be expressed in terms of spin operators in a representation independent manner, so that having solved the problem in one (e.g.  $2 \by 2$) representation  we  automatically    obtain the evolution operator for  all representations.  In order to write all the transition probabilities explicitly, one additionally requires the matrix elements of the Wigner matrix for $SU(2)$ which are readily available\cite{Hamermesh}. Using these   properties of the 
underlying Lie algebra, Hioe  solved the general $SU(2)$ LZ problem\cite{Hioe}. Since this problem  directly descends from the $2 \by 2$ case through a well-defined general procedure, we group them together and consider the general $SU(2)$ problem   trivially integrable as well. Various limits of the $SU(2)$ problem  also generate solvable LZ Hamiltonians in the same trivially integrable $2\by 2$ class.
For example, the large spin limit yields a driven harmonic oscillator with   a time dependent frequency\cite{Sinitsyn_oscillator}
\beg 
H_{o}(t) = ta^{\dag}a + g_{o}(a^{\dag} + a),
\label{osc}
\en
where $a^{\dag}$ ($a$) is the particle creation (annihilation) operator and we denote the coupling constant as $g_{o}$ to distinguish it from that  in the spin Hamiltonian.  A further limit of large occupancy transforms the oscillator into  a time dependent linear chain\cite{Sinitsyn_LC}
\beg 
H_{lc}(t) = \sum_{n = -\infty}^{\infty}[nt\ket{n}\bra{n} + (g_{lc}\ket{n}\bra{n+1} + c.c.)],
\label{lin_chain}
\en
 in Dirac bra-ket notation. Again, to separate the coupling constant from the above two cases, we have denoted it as $g_{lc}$. To illustrate the descendant generation procedure,  we     explicitly derive   the LZ transition probabilities for both these models as well as the general $SU(2)$ Hamiltonian from the known solution of the $2\by 2$ LZ problem. 
 
 One can also express the $2\by2$ LZ Hamiltonian in terms   of the $SU(1,1)$ algebra as $H(t)=tK_0-\i g K_1$, where $K_0=\frac{\sigma_z}{2}, K_1=\i\frac{\sigma_x}{2}$ and $K_2=\i\frac{\sigma_y}{2}$ are the generators of the $2$-dimensional non-unitary representation of $SU(1,1)$\cite{Conrady, Basics}.  Similar to the $SU(2)$ generalization,  we promote $K_0, K_1$, and $K_2$ to any other $SU(1,1)$ representation and solve the resulting LZ problem. For example, two well-known realizations of the $SU(1,1)$ algebra are the one mode (nonlinear oscillator) and two-mode (double oscillator) realizations. The first one leads
 to the following  LZ problem:
 \beg
 H(t)=\frac{t}{2}a^\dagger a+\frac{\tilde g}{2}\left[ (a^\dagger)^2+a^2\right].
 \label{1modeH}
 \en
And the two-mode realization results in\cite{Sinitsyn_oscillator}
\beg
H(t)=t a^\dagger a+ \tilde {g}\left(a^\dagger b^\dagger +ab\right),
\label{2modeH}
 \en
where $a^\dag$ ($a$) and $b^\dag$ ($b$) are independent bosonic creation (annihilation) operators. We obtain  transition probabilities for both these LZ problems through the same procedure as in the $SU(2)$ case using  known formulas for the Wigner-Bargmann matrix elements\cite{Bargmann, Ui1, Ui2, Ui3, Conrady}. These and other $SU(1,1)$ descendants of
the basic $2\by2$ LZ problem therefore also belong to the same trivially integrable class. We emphasize
that `trivially integrable'  only means that any polynomial commuting partner must  reduce to a combination of the Hamiltonian and identity and does not imply that the corresponding LZ problem is somehow obvious or unimportant. Equivalently, one can say that any $2\by 2$ Hamiltonian linear in a parameter belongs to the maximal integrable family consisting of itself and identity. This is similar to one dimensional Hamiltonian systems in Classical Mechanics, which are all integrable because there is one degree of freedom and one integral of motion -- the Hamiltonian itself.

This method of producing LZ solvable descendants can in principle be applied to any other solvable LZ matrix Hamiltonian\cite{Galitsky}. Consider, for example,
 the $3 \by 3$ bow-tie  Hamiltonian. It can be expressed as a linear combination of   Gell-Mann matrices from the fundamental representation of the $SU(3)$ algebra. In the same way as the $2\by 2$ LZ problem is extended to an arbitrary $SU(2)$ spin, one can   replace the  $3 \by 3$ $SU(3)$ matrices in the bow-tie Hamiltonian  with corresponding higher dimensional ones. The result is a new exactly solvable LZ problem. The general  time evolution operator  follows directly  from the  $3\by 3$ bow-tie LZ problem  similar to the $SU(2)$ case discussed above.
 Now with the help of the   formula for a general Wigner matrix element for $SU(3)$ group \cite{Prakash} one can    determine the transition probabilities for the generalized problem.  Higher dimensional equal slope, bow-tie and generalized bow-tie Hamiltonians can be similarly written as  linear combinations of generators of some $SU(L)$ algebra, where $L>3$,  and thus generalized to higher dimensions at least in principle.

The rest of the paper is organized as follows. In Section II we give a brief review of quantum integrability. We consider  an example when all the commuting partners are linear in a real parameter in more detail. In Sections III, IV, and V we address equal slope, bow-tie,  and generalized bow tie models, respectively. For each of these models we explicitly construct    commuting real-symmetric operators that are either linear (for equal slope and generalized bow-tie cases) or quadratic (for bow-tie case)  in $t$.  We take up the generalized spin and related cases and argue that they are ``trivially" integrable in Section VI.  

\section{Integrable parameter dependent matrices:  short review}
\label{Rev}
 
In classical mechanics a Hamiltonian with $n$ degrees of freedom is said to be integrable if it has $n$ (maximum possible number) nontrivial integrals of motion --   independent functions of generalized coordinates and momenta that Poisson commute with the Hamiltonian and among themselves.  Then the Hamilton-Jacobi equation is completely separable,   equations of motion are exactly solvable, and the motion is constrained to   an  invariant torus\cite{Arnold}. However, it is well known that a direct  import of this definition to the realm of quantum mechanics is fraught with severe difficulties. Main obstacles are that there are no natural well-defined notions of a nontrivial integral and of the number of degrees of freedom, see e.g.    \cite{Shastry} for further discussion. The latter issue is especially problematic for a quantum system living in a finite dimensional Hilbert space.  

A possible resolution of these difficulties was proposed in  \cite{Shastry2005, Emil1, Shastry2011, Emil2, Shastry}. It was observed that in many exactly solvable condensed matter models,
such as  1d Hubbard, XXZ, Richardson and Gaudin models, the Hamiltonian depends linearly on a certain parameter, typically an interaction or an external field strength. Let us therefore consider an abstract Hamiltonian of the form $H(u)=T+u V$, where $T$ and $V$ are $N\by N$ Hermitian matrices and $u$ is a real parameter. In the Landau-Zener problem we identify $u=t$ or $u=bt$ in the equal slope model. For generic $T$ and $V$, e.g. two randomly generated such matrices,  the only Hermitian matrix similarly linear in $u$ that commutes with $H(u)$ is $(a + bu)\mathbb{1} + cH(u)$, where $a,b$ and $c$ are real numbers and $\mathbb{1}$ is the identity matrix. We identify this as the trivial commuting partner. It turns out that the requirement that $H(u)$ has a nontrivial commuting partner is a very restrictive one, so that the set of $H(u)$ with at least one such nontrivial integral of motion  is of measure zero among Hermitian matrices of the form $T+uV$. 

A Hamiltonian matrix $H(u)=T+u V$  is said  \cite{Emil1,Emil2} to belong to a linear family of integrable matrices  if it has $n\ge 1$ nontrivial   partners linear in $u$ that commute with $H(u)$ and each other and, in addition,   have no common $u$-independent symmetry. The last condition means that there is no Hermitian $u$-independent   matrix that commutes with $H(u)$ and all its integrals of motion.  Whenever  common $u$-independent symmetries are present, the Hamiltonian and its commuting partners  can be simultaneously block-diagonalized resulting in a commuting set of smaller matrices until all such symmetry is exhausted. This definition naturally leads to a classification of linear integrable matrices according to the number $n$ of nontrivial integrals. It turns out that the maximum possible number is $n=N-2$, so $n$ ranges from 1 to $N-2$ and the $2\by 2$ case is trivial as we already commented above. 

Remarkably, from this definition alone \textit{all}  linear integrable models with $n=N-2, N-3,$  $N-4$ as well as a certain class of models with any other allowed number of nontrivial commuting partners were constructed explicitly in  \cite{Emil1,Emil2}. Moreover, all these models  turn out to be exactly solvable and exhibit Poisson level statistics and energy level crossings, even when there is only a single commuting partner. We believe this justifies the name ``integrable'' at least for these models  even   when there are only a few integrals. Apparently   fixing the parameter dependence in the Hamiltonian and its commuting partners has powerful consequences. 

As an example relevant to the present paper, consider the maximal family of commuting matrices linear in $u$, i.e. $n=N-2$. Counting the Hamiltonian itself and the identity matrix, these are $N$ mutually commuting $N\by N$ matrices of the form $H_i(u)=T_i+u V_i$, i.e. $[H_i(u), H_j(u)]=0$ for all $i, j$, and $u$.   In the shared eigenbasis of $V_i$ any maximal family can be explicitly parametrized as follows \cite{Emil1}
\beg
\ba{l}
\dis [H_{i}(u)]_{ij} = \frac{\g_{i}\g_{j}}{\ve_{i} - \ve_{j}},\;\;i \neq j,\\
\dis \\
\dis [H_{i}(u)]_{jj} = -\frac{\g_{i}^{2}}{\ve_{i} - \ve_{j}},\;\;i \neq j,\\
\dis \\
\dis [H_{i}(u)]_{ii} = u - \sum_{j \neq i}\frac{\g_{j}^{2}}{\ve_{i} - \ve_{j}},
\ea
\label{MatEl}
\en
where $\g_i,$ and $\ve_i$ are arbitrary real parameters. The above matrix can also be written in the bra-ket notation as follows:
\beg 
H_{i}(u) =  u \out{i}{i} + \sum_{j \neq i}\frac{\g_{i}\g_{j}\out{i}{j}+\g_{i}\g_{j}\out{j}{i}-\g_{j}^{2}\proj{i}-\g_{i}^{2}\proj{j} }{\ve_{i} - \ve_{j}}.  
\label{bra-ket_type-1} 
\en
There is a certain ``gauge freedom" in the choice of the parameters. In particular, we can factor out an overall scale in $\g_i$ and
apply a constant shift to all $\ve_i$, e.g. we can set one of the $\g_i$ to  1 and one of the $\ve_i$ to $0$.  A general Hamiltonian that is  a member of this commuting is an arbitrary linear combination, $H(u)=\sum_{i=1}^N d_i H_i(u)$.

 The above notion of an integrable matrix naturally generalizes to   include commuting partners with a polynomial dependence on the parameter. In the present paper we call an $N\by N$ Hermitian $H(u)=T+uV$ integrable if it has at least one nontrivial Hermitian commuting partner $I(u)$ that is a matrix polynomial in the real parameter $u$ of order   $p\le N-2$ and if $H(u)$ and $I(u)$ have no common $u$-independent symmetry. As already mentioned above, nontriviality means that $I(u)$ cannot be expressed in terms of powers of $H(u)$ no higher then $p$, i.e. $I(u)\ne\sum_{k=0}^p c_k(u) H^k(u)$, where $c_k(u)$ are scalar  polynomial functions of $u$.
 
 The restriction on the order of $I(u)$ stems from the observation that as long as the matrix $V$ is nondegenerate, commuting partners of order in $u$ higher then $N-2$ are either trivial or reduce to lower order polynomials. Let us demonstrate this in the $3\by 3$ case. The statement in this case is that a quadratic commuting matrix $I(u) = \wt{T} + u\wt{V} + u^{2}Q$ is either trivial or 
 reduces to a linear one. Indeed, $[I(u), H(u)]=0$ implies $[Q, V]=[Q, V^2]=0$, i.e. we can simultaneously diagonalize $Q, V$, and $V^2$. There are at most three linearly independent diagonal $3\by 3$ matrices  and $\mathbb{1}$, $V$, and $V^2$ are a complete set if $V$ is nondegenerate. It follows that $Q = \a V^{2} + \b V + \g \mathbb{1}$, 
 where $\a, \b, \g$ are real numbers. 
  Now consider ${I}'(u) = I(u) - \a [H(u)]^{2} - \b H(u) - \g \mathbb{1}$.  By construction the quadratic dependence cancels out in ${I}'(u)$, so it is linear in $u$ and  commutes with $H(u)$
  thus proving our statement.   Similarly, in the general $N\by N$ case commuting matrices of order $N-1$ or higher in $u$ reduce to those of order $N-2$ or lower. The restriction
  on the order in $u$  in our definition of a nontrivial integral eliminates this redundancy. Thus, a nontrivial commuting partner of degree $p$  can  exist  only for dimensions $(p + 2) \by (p + 2)$ and higher.  In particular, the $2\by 2$ case is trivial and in $3\by 3$ one can have at most linear in $u$ integrals.
  
The classification of  linear integrable matrices  discussed  above,  can  be extended to include families   with polynomial dependence on the parameter.  For example, families with both linear and quadratic matrices are labeled as $(r,s)$, where $r$   and $s$ are the numbers of nontrivial independent linear and quadratic  commuting matrices, respectively. We define $s$ so that it cannot be reduced by combining quadratic members with squares of linear ones. In this notation the maximal (having in total $N-1$ nontrivial linear commuting members)   family is designated as $(N-1,0)$  as it can be shown that there are no independent quadratic commuting partners.  And, as we will see below, the bow-tie Hamiltonian  belongs to an  $(1,N-3)$ family. This family  contains  a single linear matrix, the Hamiltonian   itself, and $(N-3)$ nontrivial quadratic commuting partners.   

Finally, note that we can take $p_i$  to be real (and nonnegative) in Hamiltonians \re{eqsl}, \re{bt}, and \re{gbt} of the equal slope, bow-tie, and generalized bow-tie models without loss of generality. The Hamiltonians then become real symmetric. Indeed, suppose $p_k = |p_k|e^{-\i \theta_k}$. Consider the following $t$-independent unitary transformation ${H}'(t) = \mathbb{U}^{\dagger}H(t)\mathbb{U}$, where 
\beg 
\mathbb{U} = \bem
e^{\i\theta_1} & & & \\
 & e^{\i\theta_2} & &  \\
 & & \ddots &  \\
 & & & e^{\i\theta_{N}} 
\enm .
\label{unitary_trans}
\en 
It can be seen that for equal-slope and bow-tie Hamiltonian, if we take $\theta_1 = 0$, the unitary transformation introduced above gets rid of the phases of the matrix elements . Similarly by choosing $\theta_1 = \theta_2 = 0$, one can make all the matrix elements  in the generalized bow-tie model real.   Since a $t$-independent unitary transformation only changes the overall phase of the wave-function $\Psi(t)$ leaving the transition probabilities intact \cite{Brundobler},  for the purpose of solving the Landau-Zener problem one can consider the real symmetric Hamiltonians without any loss of generality. Such a transformation also does not affect the commutation relations and the $t$-dependence of any polynomial commuting partners.

Moreover, assuming $p_i$ and $r_i$ to be distinct among themselves,  commuting partners for each of the above models must also be real symmetric in the same basis where the model  Hamiltonian $H(t)$   is  real symmetric. This is a consequence of the fact that, as we show below, the spectra of $H(t)$ are nondegenerate except at a finite number of  values of $t$ and of the following lemma.  A polynomial Hermitian commuting partner of a real symmetric Hamiltonian $H(t)$ with degeneracies only at a finite number of points $t=t_i$ is  real symmetric.   Indeed, $[H(t),I(t)] = 0$ guarantees both   $H(t)$ and $I(t)$ can be diagonalized in a common basis.   The  matrix $\mathbb{O}$ that diagonalizes $H(t)$ is unique and orthogonal (rather than general unitary) except maybe at a few values of $t$.    $I(t)$ is  diagonal in the same basis, so it is diagonalized by the the same orthogonal matrix $\mathbb{O}$, i.e. $\mathbb{O}I(t)\mathbb{O}^{T} = I_{d}(t)$, where $I_{d}(t)$ is a  diagonal real matrix. Now one can use the inverse orthogonal transformation to get $I(t)$ back from $I_{d}(t)$, i.e. $\mathbb{O}^{T}I_{d}(t)\mathbb{O} = I(t)$, which is real except maybe at  $t=t_i$. Thus, $I(t)$ is a matrix polynomial in $t$    that is real at infinitely many values of $t$. It follows that  $I(t)$ is real symmetric.

\section{Equal slope model}

In this section we show that the equal slope model  (\ref{eqsl}) is of the form of \eref{MatEl}. It thus belongs to the maximal linear family and has the maximum possible number of nontrivial linear commuting partners. 
Identifying $u=bt$, we rewrite the $N \by N$ equal slope Hamiltonian as follows:
\beg
H(u) =\bem
u & p_{2} & \cdots & p_{N}\\ 
p_{2}& a_{2} & \cdots & 0\\ 
\vdots & \vdots & \ddots &  \vdots\\ 
p_{N} & 0 & \cdots & a_{N}
\enm .
\label{eqslHam}
\en
On the other hand,  $i=1$ element   of the maximally commuting linear family \re{MatEl}  reads
\beg
H_{1}(u) =\bem
\dis u - \sum_{j \neq 1}\frac{\g_{j}^{2}}{\ve_{1} - \ve_{j}} & \dis \frac{\g_{1}\g_{2}}{\ve_{1} - \ve_{2}} & \dis \cdots & \dis \frac{\g_{1}\g_{n}}{\ve_{1} - \ve_{n}}\\ 
\dis \frac{\g_{1}\g_{2}}{\ve_{1} - \ve_{2}} & \dis -\frac{\g_{1}^{2}}{\ve_{1} - \ve_{2}} & \dis \cdots & \dis 0\\ 
\dis \vdots & \dis \vdots & \dis \ddots & \dis \vdots\\ 
\dis \frac{\g_{1}\g_{n}}{\ve_{1} - \ve_{n}} & \dis 0 & \dis \cdots & \dis -\frac{\g_{1}^{2}}{\ve_{1} - \ve_{n}}
\enm .
\label{h1}
\en
We see that the equal-slope Hamiltonian has the same matrix structure as $H_{1}(u)$ in Eq.~\eqref{h1}. Now all one has to do is to determine suitable values for $\g_i$ and $\ve_i$ such that matrix elements of Eq.~\eqref{h1} map to matrix elements in Eq.~\eqref{eqslHam}. Using the gauge freedom explained below Eq.~\eqref{MatEl}, we choose $\g_1 = 1$ and $\ve_1 = 0$. We expect $H(u) = H_{1}(u) + x\mathbb{1}$, i.e. 
\beg
\bem
u & p_{2} & \cdots & p_{N}\\ 
p_{2}& a_{2} & \cdots & 0\\ 
\vdots & \vdots & \ddots &  \vdots\\ 
p_{N} & 0 & \cdots & a_{N}
\enm = \bem
\dis u + x + \sum_{j \neq 1}\frac{\g_{j}^{2}}{\ve_{j}} & \dis -\frac{\g_{2}}{\ve_{2}} & \dis \cdots & \dis -\frac{\g_{N}}{\ve_{N}}\\ 
\dis -\frac{\g_{2}}{\ve_{2}}& \dis x +\frac{1}{\ve_{2}} & \dis \cdots & \dis 0\\ 
\dis \vdots & \dis \vdots & \dis \ddots & \dis \vdots\\ 
\dis -\frac{\g_{N}}{\ve_{N}} & \dis 0 & \dis \cdots & \dis x +\frac{1}{\ve_{N}}
\enm .
\label{equiv}
\en
To achieve a one-to-one correspondence between the matrix elements, we need
\begs
\bea
x + \frac{1}{\ve_i}  &=& a_i,\;  i = 2,\cdots , N,  \label{EScond1}\\
x + \sum^{n}_{j = 2}\frac{\g_{j}^{2}}{\ve_{j}}  &=&  0, \label{EScond2}\\
\frac{\g_{i}}{\ve_{i}}  &=&  -p_i, \;  i = 2,\cdots , N.
\label{EScond3}%
\eea 
\ens
 Eqs.~\eqref{EScond1} and \eqref{EScond3}  obtain
\beg
\ba{l}
\dis \ve_i = \frac{1}{a_i - x}, \; i = 2, \cdots , N,\\
\dis \\
\dis \g_i = \frac{p_i}{x - a_i}, \; i = 2, \cdots , N.
\ea
\label{EScond4}
\en
Substituting Eq.~\eqref{EScond4} into Eq.~\eqref{EScond2}, we find the following algebraic equation for $x$:
\beg 
x = \sum^{N}_{i = 2}\frac{p_i^{2}}{x - a_{i}}.
\label{EScond5}
\en
All $N$ roots of \eref{EScond5}   are real. This is seen by e.g plotting both sides of this equation  taking into account that the right hand side varies from
$-\infty$ to $\infty$ between any two consecutive poles at $x=a_i$.
 Plugging the value of $x$ into Eq.~\eqref{EScond4}, we get a set of $\g_i$ and $\ve_i$ that map equal slope Hamiltonian into one of the members of the maximally commuting family, rendering equal slope case quantum integrable. 

Note also that in the limit $t\to\infty$  the first term on the right hand side of \eref{eqsl} is negligible and  $H(t)/t$ has an apparent $N-1$-fold degeneracy.
This is interpreted as a multiple level  crossing violating the Wigner - von Neumann noncrossing rule -- one of the tell-tale signs of quantum integrability\cite{Emil1}.

\section{Bow-Tie Model}
\label{sec:bt}

In this section we consider the bow-tie model   defined by Eq.~\eqref{bt}. Ostrovsky and Nakamura explicitly constructed the solution to the Landau-Zener problem for this 
Hamiltonian\cite{Ostrovsky}. We show that this model is member of a $(1, N-3)$ integrable family that consists of a single linear (the bow-tie model itself) and $N-3$ nontrivial quadratic commuting matrices, which we explicitly construct.
 Nontrivial quadratic partners only start to exist from dimensions $4 \by 4$.  
 
Let us redefine the real parameter $t\to u$ in Eq.~\eqref{bt}, i.e. $H(u)=T+uV$, where
\beg      
T=\bem
0 & p_{2} & \cdots & p_{N}\\ 
p_{2}& 0 & \cdots & 0\\ 
\vdots & \vdots & \ddots &  \vdots\\ 
p_{N} & 0 & \cdots & 0
\enm,\quad V=\bem
0 & 0 & \cdots & 0\\ 
0& r_{2} & \cdots & 0\\ 
\vdots & \vdots & \ddots &  \vdots\\ 
0 & 0 & \cdots & r_{N}
\enm .
\label{bt_redefined}
\en

  The energy values $(E)$ of this Hamiltonian satisfy the following characteristic equation:
\beg 
E = \sum_{k = 2}^{N}\frac{p_k^{2}}{E - ur_k}
\label{bt_char_eq}
\en 
The right hand side of the above equation has poles at $ur_k$, where $k = 2, \cdots, N$ and since all the $r_k$'s are different, for $u \neq 0,\pm\infty$ all the   poles are different as well. Now by plotting the above equation and keeping in mind that the right hand side varies from $-\infty$ to $\infty$ between any two consecutive poles at $E=ur_k$, one can see that the \eref{bt_char_eq} has $N$ distinct real roots. Thus using the  lemma proven at the very end of Sec.~\ref{Rev} one expects all the polynomial commuting partners of bow-tie Hamiltonian to be real symmetric. 

Also we see from \esref{bt_char_eq} and \re{bt_redefined}   that for this model at $u = 0$ the eigenvalue $0$ is $(N-2)$-fold degenerate. Thus we have at least one level-crossing at $u = 0$  for dimensions $4 \by 4$ and higher in the absence of any $u$-independent symmetry (see below), which is   at odds with the Wigner - von Neumann noncrossing rule \cite{Lieb1970,Shastry2002}.  The smoking gun for quantum integrability according to our definition however is the existence of a nontrivial commuting partner that depends on the real parameter of the model in a polynomial fashion. We restrict ourselves to finding   such polynomial commuting partners of minimum degree.  

\subsection{Linear Commuting Partner}
\label{sec:btlinear}

Here we  show that the bow-tie Hamiltonian does not possess nontrivial linear commuting partners. We have already stated that the first nontrivial partner, which is quadratic and not linear, appears from the $4 \by 4$ case.  
 First, let us show that there is no $u$-independent symmetry, i.e. no Hermitian $\Omega\ne a \mathbb{1}$ that commutes with
$H(u)$ for all $u$. Indeed, $[\Omega, H(u)]=0$ for all $u$ means $[\Omega, T]=[\Omega, V]=0$. It follows that  $\Omega$ is diagonal in the same basis where $V$ is diagonal and $H(u)$ takes the form of Eq.~\eqref{bt_redefined}.  In this basis
the commutation relation  $[\Omega, T]=0$ obtains $p_i(\Omega_{11}-\Omega_{ii})=0$, which  implies $\Omega_{ii}=\Omega_{11}\equiv a$ and therefore $\Omega= a \mathbb{1}$.

A general linear commuting partner can be written as
\beg 
I(u) = \widetilde{T} + u\widetilde{V}.
\label{bt:partner}
\en
The existence of a linear commuting partner entails $\left[H(u), I(u)\right] = 0.$ Equating all the coefficients of $u$ to zero we get the following commutation relation:
\beg 
[T, \widetilde{V}] = [\widetilde{T},V], \quad [T, \widetilde{T}] = [V, \widetilde{V}] = 0.
\label{bt:lincomm}
\en 
We notice that if we consider a special basis where $V$ and $\widetilde{V}$ are simultaneously diagonal (we call it the ``diagonal basis''), one of the commutation relations $([V, \widetilde{V}] = 0)$ is automatically satisfied.  
    
Our task is to determine $\widetilde{T}$ and $\widetilde{V}$ from Eqs.~\eqref{bt:lincomm}. Let us denote the matrix elements as $v_k=\widetilde{V}_{kk}$ and $t_{km}=\widetilde{T}_{km}$, where $k$ and $m$ range from 1 to $N$. There are thus $N(N+1)/2+N$ unknowns.    Eqs.~\eqref{bt:lincomm} become
\begs
\bea
p_{i}(t_{ii} - t_{11}) + \sum_{m\ne i,1} p_m t_{im}  &=& 0, \\
 p_2t_{1j} - p_j t_{12}   &=& 0, \\
p_i(v_i - v_1) & = & t_{1i}r_i, \\
t_{ij}(r_i - r_j)  &=& 0,
\eea 
\label{bt:lincommexpl}%
\ens
where $i>1$ and $j>2$.
 There are  $N(N+1)/2+N-3$ linearly independent equations, i.e. 3 less then the number of unknown matrix elements.  Any $3$ variables    can be treated as arbitrary parameters and the rest of them can be expressed in terms of those arbitrary parameters. If one considers $t_{11}, t_{12}$ and $v_1$ to be the arbitrary parameters,  Eqs.~\eqref{bt:lincommexpl}  result in
\begs
\bea
t_{ii}   &=&  t_{11}, \\
t_{1i}  & = & \frac{p_i}{p_2}t_{12}, \\
v_i  &=&   \frac{r_i}{p_2}t_{12} + v_1, \\
 t_{ij} &=&  0.
\eea 
\label{bt:explsol}%
\ens 

Next, we need to check whether the linear commuting partner  given by Eq.~\eqref{bt:explsol} is nontrivial. A trivial   partner  is of the form
\beg 
I(u) = (\a u + \b)\mathbb{1} + \g H(u).
\label{bt:trivial}
\en 
If the derived commuting partner is trivial, $\a, \b$ and $\g$ in Eq.~\eqref{bt:trivial} should have a unique solution in terms of $t_{11}, t_{12}$ and $v_1$. For our case we can indeed obtain such a unique solution.  Specifically,
\beg
\a = t_{11}, \; \b = v_1, \; \g = \frac{t_{12}}{p_2}.
\label{bt:mapping_into_trivial}
\en
This means the commuting partner that we have found is a trivial one. 
Thus we conclude that the bow-tie Hamiltonian in general does not have a nontrivial linear commuting partner. 

\subsection{Quadratic Commuting Partner}
\label{sec:btquad}

In this section we construct all nontrivial quadratic commuting partners for the bow-tie Hamiltonian.   A general quadratic commuting partner can be written as
\beg      
I(u) = \widetilde{T} + u\widetilde{V} + u^{2}Q 
\label{bt:quad_partner_non_explicit}
\en
 Now in order for $I(u)$ to commute with $H(u)=T+uV$ we need
\beg 
[T, \widetilde{T}] = 0, \quad  [T, \widetilde{V}] = [\widetilde{T}, V], \quad [T, Q] = [\widetilde{V}, V], \quad [V, Q] = 0.
\label{bt:quadcomm}
\en 
Unlike in the linear case as shown in Eq.~\eqref{bt:lincomm}, the commutation relations imply that we can achieve simultaneous diagonalization of $V$ and $Q$. So for the quadratic commuting partner we designate the shared eigenbasis of $Q$ and $V$ to be the ``diagonal basis".  

Our task is to determine $I(u)$.  In the diagonal basis we only have to resolve the first three commutation relations in Eq.~\eqref{bt:quadcomm}. Let $t_{km}=\widetilde{T}_{km},  v_{km}=\widetilde{V}_{km}$, and $q_{kk}=Q_{kk}$.
Writing out the commutation relations explicitly we get the following set of linearly independent equations:
\begs
\bea
v_{ij}(r_j - r_i) &=& 0, \\
p_{i}(v_{ii} - v_{11}) + \sum_{m\ne i,1} p_m v_{im}  &=& 0, \\
v_{1i}r_i &=& p_i(q_{ii} - q_{11}),\\ 
p_2t_{1j} - p_j t_{12}   &=& 0, \\
t_{ij}(r_i - r_j) &=& p_iv_{1j} - p_jv_{1i}, \\
p_{i}(t_{ii} - t_{11}) + \sum_{m\ne i,1} p_m t_{im}  &=& 0,
\eea
\label{bt:quad_expl}%
\ens
where $i>1$ and $j>2$. It can be shown that for an $N \by N$ dimensional case, one has $N^2 + 2N$ variables whereas the commutation relations give rise to only $N^{2} + N -3$ linearly independent equations. So in general we have an under-determined system and the difference between the number of variables and the number of independent linear equations is $N+3$. Treating $t_{11}, t_{12}, v_{11}, q_{11}, \cdots, q_{NN}$ to be arbitrary parameters, from \eref{bt:quad_expl} one obtains
\begs
\bea
v_{ij} &=& 0, \\
v_{ii}  &=&  \frac{r_i}{p_2}t_{12} + v_{11}, \\
v_{1i} &=& \frac{p_i}{r_i}(q_{ii} - q_{11}), \\
t_{1j}  &=&  \frac{p_j}{p_2}t_{12}, \\
t_{ij}  &=& \frac{p_ip_j}{r_i - r_j}\left[\frac{q_{ii} - q_{11}}{r_i} - \frac{q_{jj} - q_{11}}{r_j}\right], \\
t_{ii} &=& t_{11} - \sum_{m \neq i}\left[\frac{p_m^{2}}{r_m - r_i}\left(\frac{q_{mm} - q_{11}}{r_m} - \frac{q_{ii} - q_{11}}{r_i}\right)\right]. 
\eea 
\label{bt:quad_expl_sol}%
\ens  
Note that $t_{11}$ and $v_{11}$ on the right hand side add multiples of the identity matrix to $I(u)$, while terms involving $t_{12}/p_2$ add a multiple of the Hamiltonian.    Eq.~\eqref{bt:quad_expl_sol}  in the bra-ket notation  reads
\begin{multline}
I(u) = (t_{11} + uv_{11})\mathbb{1} + \frac{t_{12}}{p_2}H(u)+u^{2}q_{11}\proj{1} 
+ \sum_{i = 2}^{N}\left\lbrace  u^{2}q_{ii} - \sum_{m \neq i}\left[\frac{p_m^{2}}{r_m - r_i}\left(\frac{q_{mm} - q_{11}}{r_m} - \frac{q_{ii} - q_{11}}{r_i}\right)\right]   \right\rbrace\proj{i}  \\ 
+ u\sum_{i = 2}^{N}    \frac{p_i}{r_i}(q_{ii} - q_{11})  (\out{1}{i} + \out{i}{1}) + \sum_{\substack{i=2 \\ i < j}}^{N}\sum_{j = 3}^{N} \frac{p_ip_j}{r_i - r_j}\left[\frac{q_{ii} - q_{11}}{r_i} - \frac{q_{jj} - q_{11}}{r_j}\right] (\out{i}{j} + \out{j}{i}).
\label{bt:quad_partner_bra-ket}
\end{multline}   
 
Similar to the linear case, we need to check whether  $I(u)$ is a trivial  or not. We remind ourselves the form of the trivial quadratic commuting partner 
\beg  
I(u) = (\a u^{2} + \b u + \g)\mathbb{1} + (\d u + \ve)H(u) + \f H^2(u).
\label{bt:quad_trivial}
\en 
In order for the solution in Eq.~\eqref{bt:quad_expl_sol} to be trivial we should be able to solve for $\a, \b, \g, \d, \ve$ and $\f$ in terms of the parameters of the system $t_{11}, t_{12}, v_{11}, q_{11}, \cdots, q_{NN}$. Since there are $N+3$ equations  and $6$ unknowns, for $N>3$ the system is overdetermined and  it is impossible to find a solution. This proves that the derived quadratic commuting partner is  nontrivial. 

In fact,  Eq.~\eqref{bt:quad_partner_bra-ket} defines an integrable family   that consists of $N$ linearly independent quadratic  real symmetric matrices and a single linear member -- $N \by N$ bow-tie Hamiltonian -- that all mutually commute.
 To get the $k$-th quadratic member of the family, $I_{k}(u)$,  we set $q_{kk}=1$ and   the rest of $q$'s to   zero.  Also subtracting $(t_{11} + uv_{11})\mathbb{1} + \frac{t_{12}}{p_2}H(u)$ and multiplying by $r_k$ for $k>1$,  we obtain
 \begs
\begin{flalign}
& I_{1}(u) = u^{2}\proj{1} - u\sum_{i = 2}^{N}\frac{1}{r_i}\left[p_i(\out{1}{i} + \out{i}{1}) + \left(\sum_{m \neq i}\frac{p_m^{2}}{r_m}\right)\proj{i}\right] + \sum_{\substack{i=2 \\ i < j}}^{N}\sum_{j = 3}^{N}\frac{p_{i}p_{j}}{r_{i}r_{j}}(\out{i}{j} + \out{j}{i}),& \\ & I_{k}(u) = u^{2}r_k\proj{k} + up_k(\out{1}{k} + \out{k}{1}) + \sum_{i \neq 1, k}\frac{p_k^{2}\proj{i} + p_i^{2}\proj{k} - p_{k}p_{i}\out{k}{i} - p_{i}p_{k}\out{i}{k}}{(r_i - r_k)}, \qquad\text{$k \geq 2$}.&  
\end{flalign} 
\label{bt:quad_family_members}%
\ens
This is the maximum possible number of quadratic commuting partners a linear Hamiltonian can have and any other such real symmetric matrix is a linear combination of $I_k(u)$. Out of $N$    commuting partners $I_k(u)$ we count $N-3$ nontrivial ones because there are three additional matrices, $u^2\mathbb{1}, uH(u)$, and $H^2(u)$, quadratic in $u$ that commute with $I_k(u)$ and the Hamiltonian. This is consistent with the counting below Eq.~\eqref{bt:quad_trivial}, $N-3=(N+3)-6$. Specifically,  we have the following three linear combinations 
of $I_k(u)$:
\beg
\sum_{k=2}^N I_k(u)=uH(u),\quad \sum_{k=2}^N r_{k}I_k(u) = H^2(u)  -  \mathbb{1}\sum_{k=2}^N p_{k}^{2}, \quad I_1 + \sum_{k=2}^N \frac{I_k(u)}{r_{k}} =u^{2}\mathbb{1}. 
\label{iden723}
\en

Note that the $3\by 3$ bow-tie Hamiltonian does not have any nontrivial polynomial commuting partners. There are no linear partners and higher order ones are always trivial in three dimensions as explained above. For example,  in this case one can express each of the quadratic partners $I_1(u)$, $I_2(u)$, and $I_3(u)$  in terms of $H(u)$, $H^2(u)$, and 
$\mathbb{1}$. The $3\by 3$ bow-tie model also does not map to a spin-1 LZ problem of the form $gS_x+tS_z$ and therefore does not reduce to the $2\by 2$ LZ Hamiltonian. It however admits the same parametrization as higher-dimensional bow-tie Hamiltonians and thus can be regarded as a trivial member of the maximal quadratic family. In the same way any
$2\by 2$ LZ Hamiltonian is a trivial member of the maximal linear family because it can be written as $d_1H_1(u)+d_2H_2(u)$ up to a multiple of identity, where $H_1(u)$ and $H_2(u)$ are given by \eref{bra-ket_type-1} for $N=2$. Similarly one can have $4\by4$ trivial members of maximal cubic commuting family etc.
 
\section{Generalized Bow-Tie Model}
\label{sec:gbt}

In this section we consider  the   generalized bow-tie model~\eqref{gbt}, the LZ problem for which was solved by Demkov  and Ostrovsky\cite{Demkov2}. We show that this model  has a single nontrivial linear commuting partner  and therefore belongs to the minimal commuting linear family.
 Redefining $t$ in \eref{gbt} as $u$, we write $H(u) = T + uV$, where
\beg      
T=\bem
\frac{\varepsilon}{2} & 0 & p_{3} & \cdots & p_{N}\\ 
0 & -\frac{\varepsilon}{2} & p_{3} & \cdots & p_{N}\\ 
p_{3} & p_{3} & 0 & \cdots & 0\\
\vdots & \vdots & \vdots & \ddots &  \vdots\\ 
p_{N} & p_{N} & 0 & \cdots & 0
\enm,\quad V=\bem
0 & 0 & 0 & \cdots & 0\\ 
0 & 0 & 0 & \cdots & 0\\ 
0& 0 & r_{3} & \cdots & 0\\ 
\vdots & \vdots & \vdots & \ddots & \vdots\\ 
0 & 0 & 0 & \cdots & r_{N}
\enm .
\label{gbt_redefined}
\en 

 Eigenvalues $E$ of this Hamiltonian  obey the following   equation:
\beg 
E = \frac{\ve^{2}}{4E} + \sum_{k = 3}^{N}\frac{p_k^{2}}{E - ur_k}.
\label{gbt_char_eq}
\en   
The right hand side of the above equation  has poles at $E=0$ and $E=ur_k$, for $k = 3, \cdots, N$. Again if all $r_k$'s are different and none of them are equal to zero, following the same argument  as that below \eref{bt_char_eq}, one can deduce that for $u \neq 0, \pm\infty$ all the roots of \eref{gbt_char_eq} are real and distinct. Once again invoking the lemma proven at the end of Sec.~\ref{Rev}, we conclude that all  polynomial commuting partners of the generalized bow-tie Hamiltonian are real symmetric in the basis  of Eq.~\eqref{gbt_redefined}.    

Considering the Hamiltonian matrix at $u = 0$,  one can show that the eigenvalue $0$ is $(N - 2)$ fold degenerate.   Therefore, for $N\ge 4$ the energy levels cross at $u = 0$   in the absence of any $u$-independent symmetry (which can be proven using the same argument as that   for the bow-tie case) in violation of the Wigner - von Neumann non-crossing rule.    The $3 \by 3$ generalized bow-tie Hamiltonian   is identical to one of the members in the $3 \by 3$ maximally commuting linear family ($H^{3}(u)$ in our notation), and so is integrable. It should be noted that for  $N=3$ a linear Hamiltonian can only have one nontrivial linear commuting partner, i.e. the maximal and minimal commuting linear families coincide. After seeing all these indications one justifiably asks whether generalized bow tie model is integrable for all dimensions, just like previously discussed equal-slope and bow-tie models. 

As was done for the equal-slope case, first we try to write the generalized bow-tie Hamiltonian as a linear combination of the family members of the maximally commuting linear family. We start by writing out the $4 \by 4$ Hamiltonian in full as
\beg      
H(u)=\bem
\frac{\varepsilon}{2} & 0 & p_{3} & p_{4}\\ 
0 & -\frac{\varepsilon}{2} & p_{3} & p_{4}\\ 
p_{3} & p_{3} & 0 & 0\\ 
p_{4} & p_{4} & 0 & 0
\enm  +  u \bem
0 & 0 & 0 & 0\\ 
0 & 0 & 0 & 0\\ 
0& 0 & r_{3} & 0\\  
0 & 0 & 0 & r_{4}
\enm  .
\label{gbt:4by4_mat}
\en
It can be shown that
\beg 
H(u) \neq r_3H_{3}(u) + r_4H_{4}(u) + x\mathbb{1},
\label{gbt_not_type_1} 
\en 
where $x$ is a real number and $H_{3}(u)$, $H_{4}(u)$ are the members of the maximally commuting  family defined in \eref{bra-ket_type-1}. This proves the   $4 \by 4$ generalized bow-tie  model does not belong to the maximally commuting family.  The same applies to $N>4$.

Now we adopt an approach similar to the one for finding out nontrivial commuting partners of the bow-tie Hamiltonian. The general linear commuting partner, if it exists, can be written as in Eq.~\eqref{bt:partner}. Denoting the matrix elements as $v_k=\widetilde{V}_{kk}$ and $t_{km}=\widetilde{T}_{km}$, where $k$ and $m$ range from 1 to $N$,  we obtain from Eq.~\eqref{bt:lincomm} 
\begs
\bea
\ve t_{12} + \sum_{m = 3}^{N}p_m(t_{2m} - t_{1m}) &=& 0, \label{gbt_1}\\
\frac{1}{2}\ve t_{1i} + p_i(t_{ii} - t_{11} - t_{12}) + \sum_{m \neq i}p_mt_{im} &=& 0, \label{gbt_2}\\
-\frac{1}{2}\ve t_{2i} + p_i(t_{ii} - t_{22} - t_{12}) + \sum_{m \neq i}p_mt_{im} &=& 0, \label{gbt_3}\\
p_j(t_{1i} + t_{2i}) - p_i(t_{1j} + t_{2j}) &=& 0, \label{gbt_4}\\
t_{ij}(r_i - r_j) &=& 0, \label{gbt_5}\\
p_i(v_i - v_1) &=& t_{1i}r_3, \label{gbt_6}\\
p_i(v_i - v_2) &=& t_{2i}r_3, \label{gbt_7}
\eea
\label{gbt:lincommexpl}%
\ens
where $i>2$ and $j>3$. Not all of the above equations are independent.   Eq.~\eqref{gbt_4} is linearly dependent on Eqs.~\eqref{gbt_2} and~\eqref{gbt_3}. Also   Eq.~\eqref{gbt_4} [via  Eqs.~\eqref{gbt_6} and~\eqref{gbt_7}] imposes the following constraints on $v_k$'s 
\beg  
v = \frac{1}{r_3}[2v_3 - (v_1 + v_2)] = \frac{1}{r_4}[2v_4 - (v_1 + v_2)] = \cdots = \frac{1}{r_N}[2v_N - (v_1 + v_2)],
\label{gbt:gen_constraint}
\en 
where we treat $v$ as an arbitrary real number. 

Eqs.~\eqref{gbt:lincommexpl} and  \eref{gbt:gen_constraint}  yield the solution for the commuting partner  
\begs
\bea
t_{12}  &=&   \frac{v_{2} - v_{1}}{\ve}\sum_{m = 3}^{N}\frac{p_m^{2}}{r_m} , \\
t_{ii}  &=&  t_{11} - \frac{\ve}{4}v + (v_{2} - v_{1})\left[-\frac{\ve}{4r_i} + \frac{1}{\ve}\sum_{m = 3}^{N}\frac{p_m^{2}}{r_m}\right], \\
t_{22} &=& -\frac{\ve}{2}v + t_{11}, \\
t_{1i} &=& \frac{p_i}{r_i}\left(\frac{vr_i}{2} + \frac{v_{2} - v_{1}}{2}\right), \\
t_{2i} &=& \frac{p_i}{r_i}\left(\frac{vr_i}{2} - \frac{v_{2} - v_{1}}{2}\right), \\
t_{ij} &=& 0, \\
v_{i} &=& \frac{r_iv}{2} +  \frac{v_{1} + v_{2}}{2}.
\eea 
\label{gbt:gen_lin_sol_expl}%
\ens
One can now collect all the matrix elements in a single equation using bra-ket notations as follows
\begin{multline}
I(u) = (t_{11} + uv_{1})\proj{1} + (t_{11} - \frac{v\ve}{2} + uv_{2})\proj{2} + \left(\frac{v_{2}-v_{1}}{\ve}\right)\left(\sum_{m = 3}^{N}\frac{p_m^{2}}{r_m}\right)(\out{1}{2} + \out{2}{1}) + \\ 
+\sum_{i = 3}^{N}\left[\frac{u}{2}(vr_i + v_{1} + v_{2}) + \left(t_{11} - \frac{v\ve}{4}\right) + (v_{2} - v_{1})\left(-\frac{\ve}{4r_i} + \frac{1}{\ve}\sum_{m = 3}^{N}\frac{p_m^{2}}{r_m}\right)   \right]\proj{i} + \\
+ \sum_{i = 3}^{N}\frac{p_i}{2r_i}\left[(vr_i + v_{2} - v_{1})(\out{1}{i} + \out{i}{1}) + (vr_i + v_{1} - v_{2})(\out{2}{i} + \out{i}{2})\right].
\label{gbt:gen_lin_sol_expl_bra-ket}
\end{multline}

 There are $(N^{2}/2 + 3N/2) - 4$ independent linear equations for $(N^{2}/2 + 3N/2)$ unknowns in \eref{gbt:lincommexpl}.  We are left with $4$ free parameters,  which we choose to be      $t_{11}, v_1, v_2$,  and $v$ in \eref{gbt:gen_lin_sol_expl}. Following the argument outlined below \eref{bt:quad_trivial},  we see that there is a single nontrivial commuting partner in \eref{gbt:gen_lin_sol_expl}. The remaining free parameters correspond to combining this partner with multiples of the Hamiltonian and the identity.   Thus the $N \by N$ generalized bow-tie Hamiltonian belongs to a  minimal commuting linear family.  We  simplify  the commuting partner by setting $v = 1, v_{2} = 1, v_{1} = 0$ and subtracting $(-\frac{v\ve}{2}+t_{11}+t_{12})\mathbb{1}$  in \eref{gbt:gen_lin_sol_expl_bra-ket}
 \begin{multline}
I(u) = \left(u + \frac{\ve}{4} - \inv{\ve}\sum_{m = 3}^{N}\frac{p_m^{2}}{r_m}\right)\proj{2} + \inv{\ve}\left(\sum_{m = 3}^{N}\frac{p_m^{2}}{r_m}\right)(\out{1}{2} + \out{2}{1}) + \sum_{i = 3}^{N}\left[\frac{u}{2}(r_i + 1) + \frac{\ve}{4}\left(1 - \inv{r_i}\right)\right]\proj{i} + \\
 + \sum_{i = 3}^{N}\frac{p_i}{2r_i}\left[(r_i + 1)(\out{1}{i} + \out{i}{1}) + (r_i - 1)(\out{2}{i} + \out{i}{2})\right].
\label{gbt:comm_partner_bra-ket}
\end{multline}
Then the general member of this commuting family is   $\alpha H(u)+\beta I(u)+(\gamma u+\delta)\mathbb{1}$.

\section{Descendants of the $2\by 2$ Landau-Zener problem}
\label{gen_spin}

As explained in the Introduction, one can take a LZ solvable matrix Hamiltonian $H(t)=T+tV$ and generalize it by expressing  $T$ and $V$ in terms of generators $G_k$ of   a suitable Lie algebra in e.g.   its lowest dimensional representation.  This generalization is not unique. For example, the $2\by 2$ LZ problem can be  expressed in terms of both $SU(2)$ and $SU(1,1)$. For many Lie algebras the time evolution operator  can also be written generally in terms of $G_k$, e.g. as $U(t)=\exp[i\sum_k \alpha_k(t) G_k]$, i.e. in a representation independent manner. We then obtain functions $\alpha_k(t)$   from the known solution for $H(t)$. Now going to another representation of the same algebra,  we get a new, higher-dimensional  LZ  problem with a known time evolution operator\cite{Galitsky}. It remains to evaluate the matrix elements of $U(t)$ in the new representation. 
It is not important for our considerations for which Lie algebras this procedure can be explicitly  carried out. It is sufficient that this  can be done  at least in some instances. Then, we  call so derived LZ solvable Hamiltonians -- descendants of the original $H(t)$. For the purpose of classifying LZ Hamiltonians according to the number of nontrivial commuting partners, it makes sense to ``factor out" this redundancy and group the entire hierarchy  of such descendants together with their lowest dimensional ancestor $H(t)$. 

  We first consider this procedure in detail for the $SU(2)$ descendants of the $2\by 2$ LZ problem.  
Solutions of LZ problems for the linear chain and the time-dependent linear oscillator\cite{Sinitsyn_LC, Sinitsyn_oscillator} defined in the Introduction follow from the general $SU(2)$ one in particular limits, so we group them in the same class as well.  Next, we represent the $2\by 2$ problem in terms of the $SU(1,1)$ algebra and solve the resulting LZ problem specializing in positive discrete representations of    $SU(1,1)$.  In particular, we derive the transition probailities for the nonlinear and double oscillator Hamiltonians given by \esref{1modeH} and \re{2modeH}, respectively.
Because, as explained in Sect.~\ref{Rev} and below \eref{iden723}, the $2\by 2$ LZ Hamiltonian is trivially integrable, we classify all its descendants as trivially integrable as well.   

\subsection{Basic  $2\by 2$ LZ problem}

First, we summarize the essential results for the $2 \by 2$ LZ  Hamiltonian  
\beg 
 H(t) = \frac{t}{2}\s_z + \frac{g}{2}\s_x .
\label{spin:2by2}
\en     
 written here as a linear combination of $SU(2)$ generators, i.e. the Pauli matrices.  The corresponding Schr\"{o}dinger equation is
\beg  
\imath\frac{\mathrm{d} }{\mathrm{d} t}\bpm 
C_{-\frac{1}{2}}^{\frac{1}{2}}(t) \\
C_{\frac{1}{2}}^{\frac{1}{2}}(t)
\epm = \frac{1}{2} \bem
 t & \dis  g \\
 \\
\dis  g & \dis - t
\enm \bpm 
C_{-\frac{1}{2}}^{\frac{1}{2}}(t) \\
C_{\frac{1}{2}}^{\frac{1}{2}}(t)
\epm .
\label{spin_sch} 
\en
Here $C_m^{j}(t)$ are the probability amplitudes for different $m$'s  that can take values $m = -j, -j + 1, \ldots, j$ . In the present case $j = 1/2$ and $m = \pm 1/2$. The solution to the above equation is
\beg 
\bpm 
C_{-\frac{1}{2}}^{\frac{1}{2}}(t) \\
C_{\frac{1}{2}}^{\frac{1}{2}}(t)
\epm = \bem
a(t) & b(t) \\
-b^{*}(t) & a^{*}(t)
\enm \bpm 
C_{-\frac{1}{2}}^{\frac{1}{2}}(-\infty) \\
C_{\frac{1}{2}}^{\frac{1}{2}}(-\infty)
\epm ,
\label{spin:sol_2by2}
\en 
where $a(t)$ and $b(t)$ satisfy the  normalization condition
\beg 
\left |a(t) \right |^{2} + \left |b(t) \right |^{2} = 1 .
\label{spin:prob_ampl}
\en
We can see that the time evolution operator  belongs to the corresponding $SU(2)$ group. It can be shown that for time evolution from $-\infty$ to $+\infty$, the asymptotic form of $a(t)$  is such that \cite{Landau, Zener, Majorana, Stuckelberg}:
\beg 
\left |a(\infty) \right |^{2} = \exp[-\frac{\pi g^{2}}{2}] .
\label{spin:2by2_Landau}
\en 
\linebreak 
Now a general $SU(2)$ matrix for $2 \by 2$ dimensions can also be  expressed in terms of the Euler angles as   \cite{Hamermesh}
\beg 
R(\a, \b, \g) \ra \bem
\dis \cos\frac{\b}{2}\exp{(\i / 2)(\a + \g)} & \dis  \sin\frac{\b}{2}\exp{(\i / 2)(\g - \a)} \\[6pt]
\dis -\sin\frac{\b}{2}\exp{(\i / 2)(\a - \g)} & \dis \cos\frac{\b}{2}\exp{-(\i / 2)(\a + \g)}
\enm .
\label{spin_euler}
\en
So, in terms of the Euler angles Eq.~\eqref{spin:2by2_Landau} can be recast in the following form:
\beg 
\left [\cos\frac{\b}{2} \right ]^{2} = \exp[-\frac{\pi g^{2}}{2}] .
\label{spin:2by2_Landau_euler}
\en
\linebreak

\subsection{$SU(2)$ descendants}

 Promoting Pauli matrices in \eref{spin:2by2} to general spin $j$ operators, we write down the general   $SU(2)$ LZ Hamiltonian as
\beg 
H_{s}(t) = gS_{x} + tS_{z},
\label{spin_NbyN}
\en 
where $S_{x}, S_{y}$ and $S_{z}$ are the generators of $SU(2)$ algebra in $N \by N$ dimensions. The time evolution operator is a rotation operator that can be written as $U(t) = 
e^{-\i\alpha(t)S_z}e^{-\i\beta(t)S_y}e^{-\i\gamma(t)S_z}\equiv \hat R(\alpha,\beta,\gamma)$, where $\alpha(t)$, $\beta(t)$, and $\gamma(t)$ are the same as in the $2 \by 2$ problem. Using this and the functional form of the Wigner matrix elements for the $SU(2)$ group\cite{Hamermesh}, Hioe obtained\cite{Hioe}  transition probabilities from one $z$-component (say $m$) to another (say ${m}'>m$) for a general spin $j=(N-1)/2$   as      
\beg 
\ba{l}
\dis P^{j}_{m \ra {m}'} = \left |D_{m{m}'}^{j}(\a, \b, \g)\right |^{2} = \left |D_{m{m}'}^{j}(\cos\frac{\b}{2}e^{(\i / 2)(\a + \g)}, \sin\frac{\b}{2}e^{(\i / 2)(\g - \a)})\right |^{2} = \\
\dis = \left[\sum_{\m}(-1)^{\m}\frac{[(j+m)!(j-m)!(j+{m}')!(j-{m}')!]^{\frac{1}{2}}}{[(j+m-\m)!(\m)!(j-{m}'-\m)!({m}'-m+\m)!]} \by (\cos\frac{\b}{2})^{2j+m-{m}'-2\m}(\sin\frac{\b}{2})^{{m}'-m+2\m} \right]^{2} ,
\ea
\label{spin_prob_m'>m}
\en  
where  $\m = 0, 1, 2, \cdots$. Similarly, for $m>{m}'$, redefining $m-{m}' + \m \to \m$, we arrive at  
\beg 
P^{j}_{m \ra {m}'}  = \left[\sum_{\m}(-1)^{\m}\frac{[(j+m)!(j-m)!(j+{m}')!(j-{m}')!]^{\frac{1}{2}}}{[(j+{m}'-\m)!(m - {m}' +\m)!(j-m-\m)!(\m)!]} \by (\cos\frac{\b}{2})^{2j+{m}'-m-2\m}(\sin\frac{\b}{2})^{m-{m}'+2\m} \right]^{2} ,
\label{spin_prob_m>m'}
\en
where again   $\m = 0, 1, 2, \cdots$. Here $\cos\frac{\b}{2}$   in terms of the coupling constant $g$ is given by Eq.~\eqref{spin:2by2_Landau_euler}. Thus explicit solution to the $N \by N$ problem follows directly from the $2 \by 2$ Landau-Zener problem. We therefore group the general spin Hamiltonian together with the  $2\by 2$ case, which is   trivially integrable.

 \subsubsection{ Time Dependent  Oscillator and Linear Chain }

 Next, we  show that time dependent  oscillator \re{osc} and linear chain \re{lin_chain} LZ problems are  special limits of the general $SU(2)$ spin case.  We map both the Schr\"{o}dinger equations  and the transition probabilities. We do the mapping in two steps. First we  obtain the   oscillator from the $SU(2)$ spin   and then  the linear chain from the oscillator.

 To achieve the mapping between the general $SU(2)$ model and the time dependent oscillator model, we use the Holestein-Primakoff formula 
\beg 
S^{+} = \sqrt{2j}\sqrt{1-\frac{a^{\dag}a}{2j}}a, \quad S^{-} = \sqrt{2j}a^{\dag}\sqrt{1-\frac{a^{\dag}a}{2j}}, \quad S_{z} = \left(j - a^{\dag}a \right). \\
\label{Holestein_Primakoff}
\en  
The above formula relates spin raising $S^+$ and lowering $S^-$ operators with oscillator destruction $a$ and creation $a^\dag$ operators.
 We take the large spin limit ($j \ra \infty$) by neglecting terms  that contain negative powers of $j$ in the Taylor expansion of Eq.~\eqref{Holestein_Primakoff}.    In this limit The  $SU(2)$ spin Hamiltonian ~\eqref{spin_NbyN}  becomes
\beg 
H(t) = t(j - a^{\dag}a) + g\sqrt{\frac{j}{2}}(a^{\dag} + a).
\en
Thus the Schr\"{o}dinger equation for the general $SU(2)$ maps to the time dependent oscillator \re{osc} when
\beg 
j \ra   \infty, \quad \sqrt{\frac{j}{2}}g \ra  g_{o}, \quad m \ra   (j - n), \quad {m}' \ra   (j - {n}'). 
\label{spin_lin_osc}
\en

 Now we consider the mapping between the time dependent oscillator and the time dependent linear chain. The Schr\"{o}dinger equation for the oscillator model can be written from \eref{osc} as
\beg    
\i \frac{\partial C_{n}}{\partial t} = g_{o}\left(\sqrt{n}C_{n-1}+\sqrt{n+1}C_{n+1}\right) + ntC_{n}, \qquad 0\leq n < \infty , 
\label{osc_sch}
\en
where $C_n$ is the $n$th component of the oscillator wave function. Similarly from \eref{lin_chain} one notices that the Schr\"{o}dinger equation for the linear chain model takes the following form:
\beg 
\i \frac{\partial C_{n}}{\partial t} = g_{lc}\left(C_{n-1}+C_{n+1}\right) + ntC_{n}, \qquad -\infty < n < \infty . 
\label{lin_ch_sch}
\en
Note that a constant shift $n\to n+a$ is equivalent to adding a multiple of the identity $at\mathbb{1}$ to the Hamiltonian without affecting the transition probability. In other words, the transition probability from state $n$ to $n'$ depends only on $n-n'$. On the other hand, in the limit
\beg 
n \ra   \infty , \qquad \frac{n - {n}'}{n} \to 0, \qquad  \sqrt{n} g_o  \ra  g_{lc}, 
\label{osc_lin_chain} 
\en
 Schr\"{o}dinger equations  \re{osc_sch} and \re{lin_ch_sch} match, so the LZ problem for the chain is a certain limit of that for the oscillator. This completes the mapping between all three stipulated Schr\"{o}dinger equations.

 Next, we show that the transition probability from state $(j - n)$ to $(j - {n}')$ for a general spin-$j$ $SU(2)$ model maps into the transition probability from state $n$ to ${n}'$ for the time dependent oscillator, and the transition probability from state $n$ to   ${n}'$ in the time dependent oscillator model  becomes that for the linear chain model in the limits given by  \esref{spin_lin_osc} and \re{osc_lin_chain}, respectively.

 First we demonstrate the mapping of the transition probability formulas for ${n}'<n$. Combining \eref{spin:2by2_Landau_euler} and \eref{spin_lin_osc}, we  relate the Euler angle and the coupling constant  for the time dependent oscillator  as 
\beg 
\left[\cos\frac{\b}{2}\right]^{2} = \exp\left(-\frac{\pi g_o^{2}}{j}\right) .
\label{Eular_osc_1}
\en  
Using the limit spelled out in Eq.~\eqref{spin_lin_osc},  we find
\beg 
\left[\cos\frac{\b}{2}\right]^{2j + {n}' - n - 2\m} \approx \exp\left(-\pi g_o^{2}\right) .
\label{Eular_osc_2}
\en
Similarly using $1 - \left[\cos\frac{\b}{2}\right]^{2} \approx \frac{\pi g_o^{2}}{j}$ we have, 
\beg 
\left[\sin\frac{\b}{2}\right]^{n - {n}'- 2\m} \approx \left(\frac{\pi g_{o}^{2}}{j}\right)^{\left(\frac{n - {n}'}{2} + \m\right)} .
\label{Eular_osc_3}
\en 
Now applying  Stirling's formula 
\beg 
\log n! \approx n \log n - n + \frac{1}{2}\log 2\pi n , 
\label{Stirling}
\en
we obtain,
\beg 
\frac{\sqrt{(2j - n)!(2j - {n}')!}}{(2j - n - \m)!} \approx (2j)^{\left(\frac{n - {n}'}{2} + \m \right)} .
\label{osc_str}
\en 
Starting from \eref{spin_prob_m'>m} for transition probability from one $z$-component (say $j-{n}$) to another (say $j-{n}'$) for a general $SU(2)$ model  and using  Eqs.~\eqref{spin_lin_osc},~\eqref{Eular_osc_2},~\eqref{Eular_osc_3}, and~\eqref{osc_str}, one arrives at the following result:
\beg 
 \underset{\underset{\sqrt{\frac{j}{2}}g \ra g_o}{j \ra \infty}}{\lim}\sqrt{P_{j-n \ra j-{n}'}^{j}} = \sqrt{P_{n \ra {n}'}^{\textrm{osc}}} = \sum_{\m} \frac{(-1)^{\m}}{\m !} \sqrt{\frac{{n}'!}{n!}}\bpm
{n}' + (n - {n}') \\
{n}' - \m
\epm
\left(2\pi g_{o}^{2}\right)^{\left(\frac{n - {n}'}{2} + \m\right)} \exp\left( -\pi g_{o}^{2}\right),
\label{osc_prob_1}
\en 
 where $P_{n \ra {n}'}^{\textrm{osc}}$ is the transition probability from  state $n$ to state ${n}'<n$    for the time dependent oscillator model.  Now, with the help  of series expansion for associated Laguerre polynomials  
\beg 
L_{n}^{\a}(x) = \sum_{k = 0}^{n}(-1)^{k}\bpm
n + \a  \\
n - k
\epm \frac{x^{k}}{k!} ,
\en 
we obtain from \eref{osc_prob_1} 
\beg 
P_{n \ra {n}'}^{\textrm{osc}} = \frac{{n}'!}{n!}\left[\exp\left(-2\pi g_{o}^{2}\right)\right]\left(2\pi g_{o}^{2}\right)^{(n - {n}')}\left[L_{{n}'}^{(n - {n}')}(2\pi g_{o}^{2})\right]^{2} .
\label{osc_prob_2}
\en 
 which is identical to the formula for transition probability  derived earlier\cite{Sinitsyn_oscillator}.  
 
 Finally,  we relate the  transition probability   from $n$ to ${n}'<n$  for the time dependent oscillator  to that   for the linear chain.  To achieve this, we use the
 well-known  formula \cite{Szego} 
\beg 
 \lim_{n'\to \infty} L_{{n}'}^{(n-{n}')}\left(\frac{2\pi g_{lc}^{2}}{n}\right) = ({n}')^{(n-{n}')}\left(2\pi g_{lc}^{2}\right)^{-(n - {n}')}J_{(n-{n}')}\left(2\sqrt{2\pi}g_{lc}\right),
\label{Laguerre_Bessel}
\en
where we have neglected terms of the order of $1/n$.
After using Eqs.~\eqref{Stirling} and ~\eqref{Laguerre_Bessel} in \eref{osc_prob_2}, we arrive at the following result:
\beg 
P_{n \ra {n}'}^{\textrm{osc}} \to P_{n \ra {n}'}^{lc} = J_{n - {n}'}^2\left(2\sqrt{2\pi}g_{lc}\right),
\label{lin_chain_prob}
\en 
where $P_{n \ra {n}'}^{lc}$ denotes the transition probability from state ${n}'$ to another state $n$ (where $n>{n}'$). \eref{lin_chain_prob} again agrees with the known result\cite{Sinitsyn_LC}.
To achieve the mapping  for $ {n}'>n$ we start from \eref{spin_prob_m>m'} and follow the same procedure  as for $ {n}'<n$. This completes the mapping between the transition probabilities.

\subsection{$SU(1,1)$ descendants}
\label{su11}
 
As mentioned above the $2 \by 2$ LZ Hamiltonian can also be  represented in terms of the generators of the $SU(1,1)$ algebra.Thus redefining $``-\i g"$ to be the new coupling constant $g_i$, \eref{spin:2by2} can be re-written as  
\beg 
H(t) = tK_0 + g_iK_1,
\label{2_by_2_SU(1,1)}
\en  
where $K_0, K_1$ and $K_2$ are the generators of the $2$-dimensional non-unitary representation of $SU(1,1)$ and can be expressed in terms of the Pauli matrices as $\frac{\s_z}{2}, \i\frac{\s_x}{2}$ and $\i\frac{\s_y}{2}$ respectively\cite{Conrady, Basics}. Since $SU(1,1)$ is a non-compact Lie group, one has to go to infinite dimensions in order to get a unitary representation of the group. Several such representations, where the generators of the group are Hermitian are known and have been classified\cite{Bargmann, Conrady, Basics, Bishop, Ui1,  Twomode, Nonlin, Multimode, Agarwal}. In this section we only consider the positive discrete representation of the group $SU(1,1)$, which is often referred to as $D_{k}^{+}$ in the literature\cite{Bargmann}. The eigenstates $(\ket{k,\m})$ are denoted by two indexes, similar to the eigenstates of $SU(2)$. For the $D_{k}^{+}$ representation of $SU(1,1)$, $k = \frac{1}{2}, 1, \frac{3}{2}, \cdots$ and $\m = k, k+1, k+2, \cdots$\cite{Bargmann, Ui1}. 

 In this section we examine some well-known realizations of the $D_{k}^{+}$ representation and consider Hamiltonians of the  form
\beg 
H(t) = tK_0 + \tilde{g}(K_+ + K_-),
\label{SU(1,1)_gen}
\en 
where $K_{\pm} = K_1 \pm \i K_2$ are the raising and lowering operators and $K_0$ is the third generator with commutation relations
\beg 
\left[K_0,K_{\pm}\right]=\pm K_{\pm}, \quad [K_+,K_-] = -2K_0.
\label{SU(1,1)_comm}
\en
The Casimir operator  in terms of $K_0$ and $K_{\pm}$  is
\beg 
K^{2} = K_0^{2}-\frac{1}{2}(K_{+}K_{-}+K_{-}K_{+}),
\label{Casimir}
\en 
which turns out to be $k(k - 1)\mathbb{1}$ for $D_{k}^{+}$.  

We determine the probability for transition from one eigenstate to another for   various realizations of the Hamiltonian in \eref{SU(1,1)_gen} with the help of known formulas for the Wigner-Bargmann matrix elements\cite{Bargmann, Ui1, Ui2, Ui3, Conrady}.     
The first example that we consider is known as the \textit{one-mode realization}\cite{Ui1, Ui3, Bishop, Basics} of the $D_{k}^{+}$ representation for $SU(1,1)$ algebra, i.e. 
\beg 
K_+ = \frac{1}{2}a^{\dagger}a^{\dagger}, \quad K_- = \frac{1}{2}aa, \quad K_0 = \frac{1}{2}\left(a^{\dagger}a + \frac{1}{2}\right),
\label{one_mode}
\en
where $a$ and $a^{\dagger}$ are boson creation and annihilation operators. Hamiltonian \re{SU(1,1)_gen} becomes the nonlinear oscillator model \re{1modeH}, where we dropped a nonessential  multiple of the identity matrix. One can check that the above realization satisfies the commutation relations  \re{SU(1,1)_comm}. We evaluate the Casimir operator  to be $-\frac{3}{16}\mathbb{1}$. Thus the only possible values of $k$  are $\frac{1}{4}$ and $\frac{3}{4}$. 

 Although Bargmann\citep{Bargmann} originally considered only  integer and half-integer values for $k$, one can extend the  $D_{k}^{+}$ representation of $SU(1,1)$ by allowing any positive real value for $k$. This leads to the irreducible unitary representation of the universal covering group $\overline{SU(1,1)}$ known as the \textit{projective representation} of the $SU(1,1)$ group. In particular, the one mode representation corresponds to a double valued projective representation of $SU(1,1)$, with $k = \frac{1}{4}, \frac{3}{4}$.    For such projective representation corresponding to  $D_{k}^{+}$, the action of raising and lowering operators on the eigenstates and the Wigner-Bargmann matrix elements remain unaltered\cite{Ui1, Ui2, Ui3}.

 We now see that for the one-mode realization  $k = \frac{1}{4}$ corresponds to the states of the oscillator with even number of particles whereas $k = \frac{3}{4}$ corresponds to the states of the oscillator with odd number of particles. Thus denoting the oscillator number operator eigenstates as $\ket{n}$ and the $D_{k}^{+}$ basis states as $\ket{k,\m}$ the above correspondence  reads as
\beg 
\left|\frac{1}{4}, N+\frac{1}{4}\right\rangle \equiv \ket{2N}, \quad \left|\frac{3}{4}, N+\frac{3}{4}\right\rangle \equiv \ket{2N + 1},
\label{one_mode_corr}
\en
where $N = 0, 1, 2, \cdots$. 

Similar to the $SU(2)$ scenario, here also a transition is only possible between the states having same $k$. In terms of the number operator states an odd (even) state can only make a transition into another odd (even) state. The transition probability can now be written using the Wigner-Bargmann matrix elements\cite{Bargmann, Ui2, Ui3} as follows 
\begin{align}
P_{\m\ra {\m}'}^{k} & = [\Theta_{\m{\m}'}(k)]^{2}|z|^{\m-{\m}'}|1-z|^{-(\m+{\m}')}[F(k-{\m}',1-{\m}'-k,1+\m-{\m}',z)]^{2} \quad (\m\geq {\m}') \nonumber \\
 & = [\Theta_{\m{\m}'}(k)]^{2}|z|^{{\m}'-\m}|1-z|^{-(\m+{\m}')}[F(k-\m,1-\m-k,1+{\m}'-\m,z)]^{2} \quad (\m\leq {\m}'),
\label{one_mode_prob_1} 
\end{align}   
where
\begin{align} 
\Theta_{\m{\m}'}(k) &= \frac{1}{(\m-{\m}')!}\left(\frac{\Gamma(\m+1-k)\Gamma(\m+k)}{\Gamma({\m}'+1-k)\Gamma({\m}'+k)}\right)^{\frac{1}{2}} \quad \text{if}\quad \m\geq {\m}',\nonumber \\
\Theta_{{\m}'\m}(k)& = (-1)^{\m-{\m}'}\Theta_{\m{\m}}(k), 
\label{one_mode_prob_2}          
\end{align}
and $F(\a,\b,\g,z)$ is a hypergeometric polynomial. In the above two equations $k$ is either equal to $\frac{1}{4}$ or $\frac{3}{4}$ for the one-mode realization  and     $\m$ and ${\m}'$  take values $k, k+1, k+2, \cdots$. Similar to \eref{spin:2by2_Landau_euler}, where we write down the relation between the Euler angle and the coupling constant for the general $SU(2)$ model, we obtain the following relation for $SU(1,1)$ algebra:
\beg 
z = 1-\exp(2\pi \tilde{g}^{2}).
\label{z_SU(1,1)}
\en
\esref{one_mode_prob_1},~\eqref{one_mode_prob_2}, and~\eqref{z_SU(1,1)} determine transition probabilities from any state to any other state in the one-mode LZ problem.
 
Next we consider the \textit{two-mode realization} of the $SU(1,1)$ algebra\cite{Ui1, Ui2, Twomode, Basics}, i.e.
\beg 
K_+ = a^{\dagger}b^{\dagger}, \quad K_- = ab, \quad K_0 = \frac{1}{2}(a^{\dagger}a + bb^{\dagger}),
\label{two_mode}
\en 
where $a(\a^{\dagger}),b(b^{\dagger})$ are the annihilation (creation) operators corresponding to two independent  oscillators. The LZ Hamiltonian \re{SU(1,1)_gen} turns into the double oscillator model \re{2modeH}, where we have added a multiple of the conserved operator $a^\dagger a-b^\dagger b$ without affecting any of the transition probabilities.  The Casimir operator \eref{Casimir} reads
\beg 
K^{2} = K_0^{2}-\frac{1}{2}(K_{+}K_{-}+K_{-}K_{+}) =\frac{1}{4}\left[(a^{\dagger}a  -b^{\dagger}b)^{2}-\mathbb{1}\right].
\label{Casimir_two_mode}
\en 
 The two mode state $\ket{n_a, n_b}$, where $n_a$ and $n_b$ are  particle numbers in $a$ and $b$  corresponds to an $SU(1,1)$ basis state $\ket{k,\m}$ with 
\beg 
k = \frac{|n_a - n_b|+1}{2}, \qquad \m = \frac{n_a+n_b+1}{2}.
\label{corr_two_mode}
\en 
Thus this is indeed a realization of the $D_{k}^{+}$ series for $k = \frac{1}{2}, 1, \frac{3}{2}, \cdots$ and $\m = k, k+1, k+2, \cdots$\cite{Bargmann, Ui1}. Therefore, the transition probabilities  for the LZ Hamiltonian \re{2modeH}  for different $k$ sectors  are determined by \esref{one_mode_prob_1},~\eqref{one_mode_prob_2} and~\eqref{z_SU(1,1)}.   For  $n_a = n_b$  these equations  yield transition probabilities that exactly coincide with those   obtained  earlier\cite{Sinitsyn_oscillator}.

In a similar manner one can also consider other realizations of the $D_{k}^{+}$ representation. Among them the Holestein-Primakoff realization \cite{Nonlin} is equivalent to the two mode one described in \eref{two_mode}. The advantage of the former realization though, is that unlike the two-mode  one it only involves one oscillator. Also the one-mode and two-mode realizations can be further generalized into multi-mode representations\cite{Agarwal, Multimode}. For all these different realizations one now can solve the LZ problem using the explicit form of the Wigner-Bargmann matrix elements spelled out in \esref{one_mode_prob_1},~\eqref{one_mode_prob_2} and~\eqref{z_SU(1,1)}.

\section{$SU(N)$ descendants of higher-dimensional solvable LZ problems}

 The procedure of generating solvable descendants of the previous Section also applies  to higher-dimensional LZ problems\cite{Galitsky}, even though the explicit evaluation of transition probabilities  can be much more cumbersome. As an example consider the $3 \by 3$ bow-tie model. Its Hamiltonian  \re{bt} can be recast as
\beg 
H(t) = p_2\lambda_1 + p_3\lambda_4 + t(a\lambda_3 + b\lambda_8),
\label{SU(3)}
\en    
where $a, b$ are suitable real numbers and $\lambda_{i}$  are the Gell-Mann matrices. This generalizes to higher dimensions via replacement of $\lambda_i$ with the corresponding matrices from other representations of $SU(3)$. The time evolution operator is a member of the $SU(3)$ group and can be parametrized e.g. as a product $\hat R_{23}(\alpha_1,\beta_1,\gamma_1)\hat R_{12}(\alpha_2,\beta_2,\alpha_2)\hat R_{23}(\alpha_3,\beta_3,\gamma_3)$ of rotation operators of the $SU(2)$ subgroups\cite{Prakash}, where the the parameters
$\alpha_i$, $\beta_i$, and $\gamma_i$ for evolution from $t=-\infty$ to $\infty$  are provided by the known solution of the $3\by 3$ Landau-Zener problem. The Wigner matrix for $SU(3)$ is much more complicated, but known\cite{Prakash}, so one can in principle determine all the transition probabilities.
Higher dimensional LZ solvable models (equal-slope, bow-tie or generalized bow-tie Hamiltonians for $N>3$ )   similarly produce hierarchies of potentially solvable descendants when
expressed as  linear combinations of the generators of an $SU(N)$ algebra.

\section{Discussion}

In this paper we studied various known exactly solvable multistate Landau-Zener problems. It turns out that they break down into two main categories. The first one consists of the equal slope, bow-tie, and generalized bow-tie models. These are genuinely  nontrivial multistate LZ Hamiltonians in that they cannot be reduced to the $2\by 2$ LZ problem. We found that a distinctive feature of these models is that they all have nontrivial essentially parameter (time) dependent commuting partners.  They are therefore integrable parameter-dependent matrices according to the definition of quantum integrability introduced in Refs.~\onlinecite{Shastry2002,Shastry2005,Emil1, Shastry2011,Emil2, Shastry}. Specifically, the equal slope Hamiltonian belongs the known maximal family of $N$ commuting $N\by N$ matrices linear in the parameter defined in Section~\ref{Rev}.  The bow-tie model  does not have any linear commuting partners. Instead it turned out to be a member of a similarly maximal, but quadratic in the parameter, family of commuting matrices, which we explicitly constructed here.  The generalized bow-tie has a single linear commuting partner and thus belongs to a minimal linear family. The last two examples are new families of integrable matrices not contained in
 Refs.~\onlinecite{Shastry2002,Shastry2005,Emil1, Shastry2011,Emil2, Shastry}. We therefore conjectured that quantum integrability in the above sense is a necessary condition for multistate LZ solvability.
 
 An important open question is whether it is also sufficient, i.e. whether the LZ problem is solvable for any  matrix Hamiltonian $H(t)$ that has nontrivial polynomial  commuting partners or only for a certain subclass of such Hamiltonians. In particular, one can ask more narrowly if the LZ problem is solvable for commuting partners of the equal slope, bow-tie, and generalized bow-tie models. Consider, for instance, the maximal commuting family of Section~\ref{Rev}. Each of the mutually commuting basic Hamiltonians $H_i(u)$ is a different equal slope model ($u\to t)$.  The question is whether an arbitrary linear combination, $H(t)=\sum_i d_i H_i(t)$, is also an exactly solvable LZ problem and if yes, how its solution relates to that for $H_i(t)$.  
 
 One can also use our results to try to identify new LZ solvable models. For example, we see that the equal slope and bow-tie models belong to maximal linear and quadratic commuting families, respectively. It is similarly possible to construct linear $H(t)$  with the maximum allowed number of qubic or quartic nontrivial commuting partners without   any of the lower order ones. Our results suggest that the LZ problem might be solvable for such $H(t)$.

The second category of exactly solvable LZ problems are those derived from the basic $2\by 2$ LZ Hamiltonian $H(t) = \frac{g}{2}\sigma_{x} + \frac{t}{2}\sigma_{z}$ or from any of the above  nontrivial multistate LZ Hamiltonians through a Lie algebraic procedure explained in the Introduction and Section~\ref{gen_spin}. In this way the solution for the general $SU(2)$ LZ Hamiltonian $H_s(t) = gS_{x} + tS_{z}$  follows from that for the $2\by 2$  LZ problem. Various (e.g. large spin) limits of the general $SU(2)$ produce further exactly solvable LZ models. Similarly,   an $SU(1,1)$ generalization of the $2\by 2$ LZ Hamiltonian produces the nonlinear and double oscillator models \re{1modeH} and \re{2modeH} together with a complete solution of the corresponding LZ problems. We  included  such descendant LZ problems  in the same integrability class as  the original ancestor Hamiltonian.  Interestingly,  there is a somewhat similar method of generating higher dimensional commuting  matrix families from lower dimensional ones\cite{Shastry2011}. Consider e.g. two commuting $N\by N$ matrices $H(u)$ and $ I(u)$   and associated operators $\hat H=H_{ij}(u)a_i^\dagger a_j$ and $ \hat I(u)=I_{ij}(u)a_i^\dagger a_j$, where $a_i$ are fermionic or bosonic annihilation operators. The commutation of $H(u)$ and $I(u)$ implies that $\hat H(u)$ and $\hat I(u)$ also commute. Since the total particle number $n_p$ is conserved, $\hat H(u)$ and $\hat I(u)$ are block-diagonal, each block corresponding to a particular value of $n_p$. The $n_p=1$ block returns our original matrices $H(u)$ and $I(u)$, while $n_p>1$ blocks produce higher dimensional commuting matrices. It seems to make sense however to group these higher dimensional descendants together with the two original matrices in the same integrability class.

\begin{acknowledgments}

We thank B. S. Shastry for stimulating our interest in the multistate LZ problem and pointing out Ref.~\onlinecite{Brundobler}. We  also thank   him and V. Galitski
for helpful discussions.
This work was supported in part by the David  and Lucile Packard Foundation.

\end{acknowledgments}

\bibliography{Bibliography}

\end{document}